\documentclass[reprint,preprintnumbers,amsmath,amssymb,nofootinbib,aps,prx]{revtex4-1}

\usepackage{slashed}
\usepackage{amsmath}
\usepackage{amssymb}
\usepackage{amsthm}
\usepackage{hyperref}
\usepackage{indentfirst}
\usepackage{psfrag}
\usepackage{graphicx}
\usepackage[utf8]{inputenc}

\newcommand{\M}{\mathcal{M}}

\newcommand{\A}{\mathcal{A}}
\newcommand{\La}{\mathcal{L}}

\hypersetup{colorlinks=true, linkcolor=blue, urlcolor=blue, citecolor=blue, linktocpage=true}

\begin{document}

\title{Solving puzzles of spontaneously broken spacetime symmetries}

\author{I. Kharuk$^{1,3}$}
\email{ivan.kharuk@phystech.edu}
\author{A. Shkerin$^{2,3}$}
\email{andrey.shkerin@epfl.ch}

\affiliation{$^1$Moscow Institute of Physics and Technology,
Institutsky lane 9,  Dolgoprudny, Moscow region, 141700, Russia}
\affiliation{$^2$Institute of Physics, Ecole Polytechnique F\'ed\'erale de Lausanne,  CH-1015 Lausanne, Switzerland}
\affiliation{$^3$Institute for Nuclear Research of the Russian Academy of Sciences, 60th October Anniversary Prospect, 7a, Moscow, 117312, Russia}

\preprint{INR-TH-2018-002}

\begin{abstract}
We establish a classical analog of the Nambu--Goldstone theorem for spontaneous breaking of spacetime symmetries. It provides a counting rule for independent Nambu--Goldstone fields and states which of them are gapped. We demonstrate that only those symmetry group generators give rise to independent Nambu--Goldstone fields that act nontrivially on a vacuum at the origin of coordinates. Other generators give rise to auxiliary fields that must be excluded from a theory by the means of inverse Higgs constraints. The physical meaning of the inverse Higgs phenomenon and an application of our results to theories of massive gravity are discussed.
\end{abstract}

\maketitle

%%%%%%%%%%%%%%%%%%%%%%%%%%%%%%%%%%%%%%%%%%%
\section{Introduction}
%%%%%%%%%%%%%%%%%%%%%%%%%%%%%%%%%%%%%%%%%%%
\label{sec:1}

Providing an analog of the Nambu--Goldstone theorem for spontaneous symmetry breaking (SSB) of spacetime symmetries is still a challenge. The reason lies in the fact that theories undergoing such SSB behave qualitatively different from systems undergoing SSB of internal symmetries. This difference comes in two aspects. First, the degrees of freedom (DoF) associated with the action of broken generators on a vacuum are not necessarily independent \cite{Ivanov:1975zq,Low:2001bw}. For example, all possible fluctuations of a scalar domain wall background can be obtained by the action of broken Lorentz transformations on it, as well as by the action of broken translation generators \cite{Low:2001bw}. Recently, this phenomenon was investigated in detail in \cite{Endlich:2013vfa,Brauner:2014aha,Watanabe:2014fva}. One major outcome of those studies was the understanding that a possible redundancy in associating a Nambu--Goldstone field (NGF) to each broken generator is closely related to a spacetime group representation an order parameter belongs to. Namely, depending on this representation, there may exist nontrivial simultaneous transformations of NGF that yet describe the same fluctuation of the vacuum \cite{Low:2001bw,Nicolis:2013sga}. Consequently, these transformations could be considered as a special sort of gauge freedom \cite{Nicolis:2013sga}, which yields some of the NGF redundant. 

The second feature of SSB of spacetime symmetries is that some of the NGF can be gapped \cite{Nicolis:2013sga}.\footnote{By NGF we understand modes associated with the action of broken generators on the vacuum. \textit{A priori}, there is no guarantee that such modes are massless.} This peculiarity was also observed in \cite{Endlich:2013vfa} from the perspective of IR theories --- it was shown that, in order to realize a particular symmetry group linearly, one must introduce massive fields that are not radial modes. Unlike pseudo--NGF, whose mass originates from explicit symmetry breaking, the gappness of such NGF is an inherent property of SSB itself. To separate these two mechanisms, we adopt the notion ``massive NGF'' (mNGF) to refer to the gapped NGF appearing in the latter case.\footnote{In \cite{Watanabe:2012hr,Watanabe:2013iia,Watanabe:2014fva} the mixture of the two mechanisms was studied. In this paper, we limit the discussion to the case of mNGF.}

Both issues outlined above stem from the question of how many NGF must be introduced in order to realize an SSB pattern in a given dynamical system. For SSB of internal symmetries, this question is resolved by the Goldstone theorem, which prescribes to assign one NGF to each broken generator. However, for spontaneous breakdown of spacetime symmetries the general answer is unknown due to possible redundancies among NGF. On the one hand, it is known that all NGF on which broken generators are realized nonlinearly can be obtained by following so--called inverse Higgs phenomenon \cite{Ivanov:1975zq}. On the other hand, the studies carried out in \cite{Endlich:2013vfa,Nicolis:2013sga} show that effective theories may necessarily include massive nonradial modes. The latter can be redefined to transform linearly under the action of the full symmetry group and can be integrated out at low energies. Hence, they do not represent NGF in the conventional sense \cite{Coleman:1969sm}. Nevertheless, their presence may necessarily follow from the SSB pattern \cite{Endlich:2013vfa}, and, moreover, their effective Lagrangian can be fully reproduced within the coset space technique (CST). Because of these two observation, we believe it is reasonable to consider them as NGF.

The aim of this paper is to establish the general rule for counting all independent NGF. Although this question was adressed in literature for particular spacetime groups \cite{Low:2001bw,Nicolis:2013sga,Watanabe:2013iia}, to the best of author's knowledge no general criterion was provided so far. We show that the full set of NGF is obtained by assigning one NGF to each generator acting nontrivially on the vacuum at the origin.\footnote{We assume that a theory is defined on some homogeneous space $ G/H $ of the symmetry group $G$. Then, by definition, the origin is a stable point of $ H $, see Appendix \ref{sec:a} for more details.} The NGF on which one may impose inverse Higgs constraints (IHC) can be redefined to transform linearly under the action of the full symmetry group \cite{Ivanov:1975zq,Brauner:2014aha} and represent massive nonradial modes noticed in \cite{Endlich:2013vfa}. 
For the remaining broken generators, one should introduce auxiliary NGF and impose IHC on them. This generalizes the known results on this topic and provides a simple criterion for identifying redundant NGF. In particular, this implies that the knowledge of an SSB pattern and a representation of fields with nonzero vacuum expectation value uniquely fixes the number of NGF. We also clarify the physical interpretation of the procedure of eliminating auxiliary NGF via inverse Higgs phenomenon and show that they can always be expressed in terms of the true NGF.

Our results are complementary to those of \cite{Nielsen:1975hm,Watanabe:2013iia,Watanabe:2014fva}. There, the question of when \textit{independent} NGF form canonically conjugated pairs was studied, while the present paper concerns with the question when the NGF should be introduced in the first place. Both problems result in the reduction of the amount of DoF, but the underlying physics is different. We would also like to note the following difference between our work and \cite{Watanabe:2013uya}. In \cite{Watanabe:2013uya}, the mass of NGF results from an explicit symmetry breaking, while our mNGF acquire mass via spontaneous symmetry breaking mechanism.

The paper is organized as follows. In Sec. \ref{sec:2}, we consider two theories undergoing SSB of spacetime symmetries. The first one is aimed to demonstrate how mNGF appear in the process of SSB. The second one includes the use of inverse Higgs phenomenon and illustrates its physical meaning. Section \ref{sec:3} covers major consequences of our analysis, including the classical analog of the Goldstone theorem. Therein we also make contact with other works in the field and comment on the relevance of our results to theories of massive gravity. Finally, Sec. \ref{sec:4} contains a brief summary of the results and concludes.

%%%%%%%%%%%%%%%%%%%%%%%%%%%%%%%%%%%%
\section{Preliminary examples}
%%%%%%%%%%%%%%%%%%%%%%%%%%%%%%%%%%%%
\label{sec:2}

For simplicity, in this section we work in the Euclidean space, which allows us to disregard the question of stability of solutions and focus on their symmetry aspects.

%%%%%%%%%%%%%%%%%%%%%%%%%%%%%%%%%%%%
\subsection{Massive NGF}
%%%%%%%%%%%%%%%%%%%%%%%%%%%%%%%%%%%%
\label{sec:2-1}

We would like to start by providing an example of a theory whose effective Lagrangian includes massive NGF. It is defined on $d$--dimensional Euclidean space and consists of two fields charged under the spatial and internal Poincare groups, $ISO(d)_{ST}$ and $ISO(d)_{int}$ accordingly. The first field is a $d$--component scalar $ \varphi_a(x) $ belonging to the co--fundamental representation of $ SO(d)_{int} $ and on which the internal translations act as shifts,
\begin{equation}
\varphi^a(x) \rightarrow \Omega^a_b \varphi^b(x) + c^a \,, ~~ \Omega^a_b \in SO(d)_{int}\,,~~ c^a \in \mathbb{R}\;.
\end{equation}
The second field $ V^i_a(x) $ is a vector and co--vector with respect to the spatial and internal Poincare groups accordingly, with the internal translations realized trivially,\footnote{The fact that $ V^i_a $ transforms trivially under the internal translations allows us to use the terms containing $ V^i_a $ without derivatives (unlike $ \varphi^a $) in the Lagrangian. }
\begin{equation}
V^i_a(x) \rightarrow \Omega_a^b \Lambda^i_j V^j_b(x) \;,
\end{equation}
where $ \Omega_a^b \in SO(d)_{int}\,, ~~ \Lambda^i_j \in SO(d)_{ST} $.
The Lagrangian of the theory reads
\begin{equation}
\begin{aligned} \label{Largr_mNGF}
\mathcal{L}=-\frac{1}{2}(\partial_i \varphi^a)^2 + \frac{1}{4} \big( \partial_{[i} V^a_{j]} \big)^2 + \varkappa V^i_a \partial_i \varphi^a + \\ + \frac{\lambda}{4d} \left( V^i_a V^a_i - dM_V^2 \right)^2 \;,
\end{aligned}
\end{equation}
where $ \varkappa, ~ \lambda $, and $ M_V $ are some positive constants and square brackets stand for antisymmetrization in the corresponding indices. We are interested in the background solutions with the following asymptotics at infinity,
\begin{equation}\label{AsAtInf}
\varphi^a(x) \sim \mu^2 x^a\,, ~~~ V^i_a(x) \sim \text{const}\,, ~~~~ \text{when} ~~~~ x^a \rightarrow \infty \;,
\end{equation}
where $ \mu $ is some constant with unit mass dimension. Assuming $ \lambda M^2_V > \varkappa^2 $, the solution fulfilling this requirement reads\footnote{Note that the value of $\mu$ is not fixed by the Ansatz describing the asymptotic behaviour at infinity, hence the variation of $ \varphi^a $ on the boundary is nonzero. Requiring the boundary term arising from varying Lagrangian (\ref{Largr_mNGF}) with respect to $ \varphi^a $ to vanish then fixes $\mu$ as in Eq. (\ref{VEV_mNGF}). }
\begin{equation} \label{VEV_mNGF}
\begin{gathered}
\varphi^a = \mu^2 x^a\,, ~~ V^i_a = M \delta^i_a \,, \\ 
M = \sqrt{M^2_V - \frac{\varkappa^2}{\lambda}}  \,, ~~ \mu^2 = \varkappa M \;.
\end{gathered}
\end{equation}

We now study fluctuations on top of this background. To identify NGF, we first determine the broken symmetry generators. The SSB pattern corresponding to solution (\ref{VEV_mNGF}) is 
\begin{equation} \label{Pattern_mNGF}
ISO(d)_{ST} \times ISO(d)_{int} \rightarrow ISO(d)_V \;,
\end{equation} 
where $ ISO(d)_V $ is a semidirect product of $ P_V^i = P^i_{ST} - \mu^2 P_{int}^i $ and $ SO(d)_V$ --- diagonal subgroup of $SO(d)_{ST} \times SO(d)_{int} $. Because only $ SO(d)_V $ is unbroken, in the spontaneously broken phase we do not distinguish between spatial and internal indices. Thus, the NGF are given by translations of $ \varphi^a $ and simultaneous internal rotations of $ \varphi^a $ and $ V^i_a $. Note, however, that an arbitrary rotation of $ \varphi^a $ can be expressed in terms of its (internal) translation,
\begin{equation}
\begin{gathered} \label{Multi_NGF}
e^{i\bar{M}_{cd} \omega^{cd}} \varphi^a = \mu^2 x^a + \mu^2 ( \Omega^a_b - \delta^a_b )x^b  = e^{i\bar{P}_c \psi^c} \varphi^a \,, \\ \psi^a = \mu^2 ( \Omega^a_b - \delta^a_b )x^b \;, 
\end{gathered}
\end{equation}
where $ \bar{P}_a $ and $ \bar{M}_{ab} $ are generators of internal translations and rotations accordingly. Then, to simplify the calculations and ensure that the coordinates do not enter the effective Lagrangian, we parametrize the fluctuations of the fields as 
\begin{equation} \label{NGF_Param}
\begin{gathered}
\varphi^a(x) = \mu^2 x^a + \psi^a(x) \,, ~~~ V^i_a(x) = \Omega^i_a(x) M \,, \\ \Omega^i_a = \delta^i_a + \omega^i_{a} - \frac{1}{2} \omega^i_{ b} \omega^b_{ a} + ... \;,
\end{gathered}
\end{equation}
where dots stay for higher order terms in $ \omega^i_{ a} $, and $\psi^a$ and $\Omega^a_b$ are independent. Substituting this into Eq. (\ref{Largr_mNGF}) and restricting ourselves to the second order in $ \omega^i_{ a} $,  we obtain the effective Lagrangian,
\begin{equation} \label{EffLagr}
\mathcal{L}_{\psi,A} = -\frac{1}{2} (\partial_i \psi^a)^2 + \frac{1}{4} ( \partial_{[i} A^a_{j]} )^2 - \frac{1}{2} \varkappa^2 A^i_j A^j_i + \varkappa A^i_a \partial_i \psi^a  \;, 
\end{equation}
where we have switched to the canonically normalized filed $ A^i_a = M \omega^i_{a} $. To rewrite this Lagrangian in a more convenient form, we redefine DoF as follows,
\begin{equation} \label{Redefine_Direct_Ihc}
\varkappa A_{ij} = \partial_i\psi_j - \partial_j\psi_i + \varkappa\tilde{A}_{ij} \;.
\end{equation}
Note that, as it can be verified from Eq. (\ref{NGF_Param}), the field $ \tilde{A}^i_j $ transforms linearly under the action of the full symmetry group. As we will see at the end of this section, redefinition of DoF (\ref{Redefine_Direct_Ihc}) corresponds to extracting the ``inverse Higgs part'' of $ A^i_j $ in the coset space framework. In the new variables the Lagrangian takes the form
\begin{equation}
\mathcal{L}_{\psi,\tilde{A}} = -\frac{1}{4}\left( (\partial_i\psi^a)^2 + (\partial_a\psi^a)^2 \right) - \frac{\varkappa^2}{2}\tilde{A}^i_j \tilde{A}^j_i + \frac{1}{4}( \partial_{[i}\tilde{A}^k_{j]} )^2 \;. 
\end{equation}
The peculiarity of this Lagrangian is that besides $ \psi^a $, it also contains the massive antisymmetric vector field $ \tilde{A}^i_a $,\footnote{The mass of the original field $ A^i_a $ comes from the interaction term between $ \varphi^a $ and $ V^i_a $ in Eq. (\ref{Largr_mNGF}). In particular, because in Eq. (\ref{NGF_Param}) we rotate $ V^i_a $ but not $ \varphi^a $, the linear in the background values of the fields term does contribute to the effective Lagrangian. It should be noticed that if we choose to rotate both fields, all terms in the effective Lagrangian will contain at least one derivative of the NGF. However, because the background solution depends on the coordinates, the effective Lagrangian will depend on them as well. Then, as it can be verified, the term coming from the interaction part will, in fact, represent a mass term for $ A^i_a $.} which is not a radial mode. Let us also note here that in theory under consideration the mass of radial modes is of order $ M $.

Let us discuss the physical nature of the field $ \tilde{A}^i_j $. As long as by the NGF one understands a field transforming non-linearly under the group action, $ \tilde{A}^i_j $ is not one of them. However, we believe that it is more appropriate to define NGF as modes associated with independent fluctuations of the vacuum. Such definition is physically justified as it shows which fields must be present in the theory to realize a given SSB pattern dynamically. In this sense, $ \tilde{A}^i_j $ is indeed a NGF, similar to ones studied in \cite{Endlich:2013vfa}.

Note also that $ \tilde{A}^i_j $ can be distinguished from matter fields, possibly present in the theory, by the fact that its dynamics is considerably restricted by the symmetry breaking pattern. Indeed, the Lagrangian of the matter fields is only subject to the general requirement of the invariance under the group action. On contrary, the Lagrangian for $ \tilde{A}^i_j $ necessarily includes the mass term, and, further, the kinetic term for $ \tilde{A}^i_j $ must sum up to
\begin{equation}
\mathcal{L}_{kin} = \frac{1}{4} \left( \partial_{[i} \Omega_{j]}^a(\tilde{A}^k_l/M) \right)^2 \;.
\end{equation}
These features of the effective theory allow to recognize $ \tilde{A}^i_j $ as a NGF.

At energy scales much below $ \varkappa $, one can integrate the field $\tilde{ A}^i_a $ out. The resulting Lagrangian reads
\begin{equation} \label{IntOut}
\mathcal{L}_{\psi} = - \frac{1}{4} \left( (\partial_i \psi^a)^2+ (\partial_a \psi^a)^2 \right) \;.
\end{equation}
Thus, by the direct expansion of Lagrangian (\ref{Largr_mNGF}) on top of background (\ref{VEV_mNGF}), we obtained the Goldstone sector of the effective theory, Eq. (\ref{EffLagr}), containing the gapped mode, and the low energy limit of this sector, Eq. (\ref{IntOut}), describing the massless modes. Note that the mass $\varkappa$ of the vector field is not fixed by the symmetry breaking pattern. Moreover, one can choose it to be of the order of or higher than the strong coupling scale $ M $,\footnote{We estimate the strong coupling scale as the scale at which the energy of fluctuations becomes comparable with the energy of the background solution. This estimate coincides with the one following from the analysis of the suppression of higher dimensional derivative terms in the coset space framework.} in which case the dynamics of mNGF can be neglected in the whole range of validity of the effective theory. Still, it is important to know about it, since the UV completion of the effective theory necessarily includes this field. Thus, we see that the field $ \tilde{A}^i_j $ plays the role of massive non-radial modes observed in \cite{Endlich:2013vfa}.

Let us now study SSB pattern (\ref{Pattern_mNGF}) from the CST perspective. Following the standard rules \cite{Ogievetsky1974}, we consider the coset space
\begin{equation} \label{ParamISO}
g_H = e^{iP_{V i} x^i} e^{i\bar{P}_a\psi'^a} e^{\frac{i}{2}\bar{M}_{ab}\omega'^{ab}} \;,
\end{equation}
where $ \psi'^a $ and $ \omega'^{ab} $ are the NGF for the broken internal translations and rotations correspondingly and the unbroken combination of translations\footnote{By definition, in this section we call a generator unbroken if its action on the background solution is trivial. This allows us to distinguish between nonlinearly realized generators and broken ones.} gives rise to the coordinates in the broken phase of the theory. As a first step, we would like to verify that the NGF introduced in this way coincide (up to constant multipliers) with their counterparts appearing in the direct expansion, Eq. (\ref{NGF_Param}).\footnote{Note that this check is nontrivial since one can use various parametrizations of the fluctuations as well as of the coset space \cite{Klein:2017npd}.} For this purpose, we study their transformation properties under the action of the symmetry generators. By acting by $ e^{i\bar{P}_a q^a} $ and $ e^{i\bar{M}_{ab} \alpha^{ab} } $ on coset space (\ref{ParamISO}) with constant parameters $ q^a $ and $ \alpha^{ab} $, we find the transformation law of $ \psi'^a $ and $ \omega'^{ab} $ to be
\begin{equation} \label{Transform_Properties}
\begin{gathered}
e^{i\bar{P}_a q^a}: ~~ \psi'^a \rightarrow \psi'^a + q^a \,, ~~ \omega'^{ab} \rightarrow \omega'^{ab} \;, \\
e^{i\bar{M}_{ab} \alpha^{ab} }: ~~~~ \psi'^a \rightarrow \Omega(\alpha)^a_b \psi'^b + \left( \Omega(\alpha)^a_b - \delta^a_b \right) x^b \,, \\ \omega'^{ab} \rightarrow \omega'^{ab} + \alpha^{ab} + ... \;,
\end{gathered}
\end{equation}
where dots stand for higher order terms. As it can be verified, $ \psi^a $ and $ \omega^i_{a} $ defined in Eq. (\ref{NGF_Param}) have the same transformation properties, and, hence, represent the same DoF.

Further, to obtain the ingredients for the construction of the effective theory, we calculate the Maurer--Cartan forms for coset space (\ref{ParamISO}), 
\begin{equation}
g_Hdg_H^{-1}= i\omega^i_{P_V}P_{V i} + i\omega^a_{\bar{P}}\bar{P}_a + i\omega^{ab}_{\bar{M}}\bar{M}_{ab} \; .
\end{equation}
Up to the linear order they are given by
\begin{equation} \label{MCF}
\omega_{P_V}^i = dx^i \, , ~~
\omega_{\bar{P}}^a = d\psi'^a - \mu^2\omega'^a_{ b}dx^b  \,, ~~ 
\omega_{\bar{M}}^{\mu a} = d\omega'^{\mu a} \; .
\end{equation}
The tetrads, the metric and the covariant derivatives of the NGF can be readily read out from Eq. (\ref{MCF}),
\begin{equation}
\begin{gathered}
e^i_j = \delta^i_j \, , ~~~~ g_{ij} = e^k_i e^l_j \delta_{kl} = \delta_{ij} \, , \\
D_i \psi'^a = \partial_i \psi'^a - \mu^2\omega'^a_{ i} \, , ~~~~
D_i \omega'^{ab} = \partial_i \omega'^{ab} \; . 
\end{gathered}
\end{equation} 
Then, the part of the effective Lagrangian, containing the kinetic term for $ \psi'^a $, the mass term for $ \omega'^i_a $, and their interaction, is reproduced in the CST as
\begin{equation}
- \frac{1}{2}(D_i \psi'^a)^2 = -\frac{1}{2}(\partial_i \psi'^a)^2 - \frac{1}{2}\varkappa^2 A_a^i A_a^i + \varkappa A^i_a \partial_i \psi'^a\;,
\end{equation}
where we have switched to the canonically normalized field $ A^i_a = M \omega'^i_a $. This coincides with the corresponding part of Eq. (\ref{EffLagr}) upon the identification $\psi'^a=\psi^a$, $\omega'^{ab}=\omega^{ab}$, which will be assumed from now on. Finally, the kinetic term for $ A^i_a $ can be reproduced straightforwardly, since the covariant derivative of $ \omega^{ab} $ coincides with the usual partial derivative. Thus, we see that Lagrangian (\ref{EffLagr}) is fully reproduced within the coset space approach. In particular, one can integrate the field $ A^i_j $ out and reproduce the low energy Lagrangian (\ref{IntOut}).

Let us now discuss the inverse Higgs phenomenon. For the case under consideration, IHC read,
\begin{equation} \label{IHCMG}
D_{[i} \psi_{j]} = 0 \,: ~~~ \partial_i \psi_a - \partial_a \psi_i = \varkappa A_{ia} \; .
\end{equation}
Because of the symmetry restrictions, in the absence of matter fields this gives the expression for $ A^i_j $ in terms of $ \psi^a $ one would have obtained by integrating $ A^i_j $ out \cite{Nicolis:2013sga,delacretaz2014re}. Hence, by using the left covariant derivatives one can reproduce the low energy limit of the theory. In particular, low energy Lagrangian (\ref{IntOut}) is reproduced as
\begin{equation}
\mathcal{L}_{\psi} = -\frac{1}{8} \left( D_{\lbrace i} \psi_{j\rbrace} \right)^2 \;.
\end{equation} 
Note, however, that instead of imposing IHC one can introduce a new variable $ \tilde{A}^i_j$ according to
\begin{equation} \label{IHC_as_DoF_change}
D_{[i} \psi_{j]} = \tilde{A}^i_j \;.
\end{equation}
Since the l.h.s. of this equation contains $A^i_j$ without derivatives, it represents a valid change of variables. It allows us to switch from the field $ A^i_j $, which transforms nonhomogeneously under the action of the symmetry group, in favor of the field $ \tilde{A}^i_j $ transforming linearly under all symmetries \cite{Brauner:2014aha}. Such redefinition of DoF corresponds to extracting the ``inverse Higgs part'' from $ A^i_j $ and it is precisely the change of variable we made before, Eq. (\ref{Redefine_Direct_Ihc}). The redefinition can always be performed and, by itself, does not reduce the amount of NGF. Consequently, when dealing with SSB of spacetime symmetries, the right question to ask is not whether one should impose IHC or not, but whether $ \tilde{A}^i_j $ will be present in the theory or not. In the studied example the answer to this question is positive, since $ \tilde{A}^i_j $ is needed to describe $ V^i_a $'s fluctuations. In the next section we consider the theory where this is not the case and explore the inverse Higgs phenomenon from one more perspective.

%%%%%%%%%%%%%%%%%%%%%%%%%%%%%%%%%%%%
\subsection{Inverse Higgs phenomenon and redundant fields}
%%%%%%%%%%%%%%%%%%%%%%%%%%%%%%%%%%%%
\label{sec:2-2}

%%%%%%%%%%%%%%%%%%%%%%%%%%%%%%%%%%%%
\subsubsection{The model}
%%%%%%%%%%%%%%%%%%%%%%%%%%%%%%%%%%%%
\label{sec:2-2-1}

To clarify the physical meaning of IHC in cases when some of the NGF are redundant, we track the way they appear during the direct calculation of the effective Lagrangian. Consider the theory which, besides the fields $ \varphi^a $ and $ V^i_a $ introduced in the previous section, contains a scalar field $ \theta $, with the Lagrangian
\begin{equation} \label{Ill_Ihc}
\mathcal{L}= -\frac{1}{2}( \Box \varphi^a)^2 - \frac{1}{2} ( \partial_i \theta )^2 + \frac{1}{4} \big( \partial_{[i} V^a_{j]} \big)^2 + \lambda \theta V^i_a \partial_i \varphi^a  \;,
\end{equation}
where $ \Box = \partial_i \partial^i $ and $\lambda$ is a constant. Such theory has the same symmetries as in the example of Sec. \ref{sec:2-1}, but the presence of $ \theta $ and the box operator ensure that
\begin{equation} \label{Sol_Ill_Ihc}
\varphi^a = \mu^2 x^a \,, ~~~~ \theta = 0\,, ~~~~ V^i_a = 0
\end{equation} 
is a solution of equation of motion with arbitrary $ \mu^2 $. Clearly, background (\ref{Sol_Ill_Ihc}) invokes the same SSB pattern as in the previous example, (\ref{Pattern_mNGF}). However, now the NGF for the broken Lorentz generators are redundant, since the NGF sector of the effective theory contains only $ d $ DoF describing the fluctuations of $ \varphi^a $. 
 
The effective Lagrangian for this theory can be found to be
\begin{equation} \label{Eff_Lagr_S2}
\begin{aligned}
\mathcal{L}_{\psi} =  -\frac{1}{2}( \Box \psi^a )^2 - \frac{1}{2} ( \partial_i \theta )^2 + \frac{1}{4} \big( \partial_{[i} V^a_{j]} \big)^2 + \\
+ \lambda \theta V^i_a ( \mu^2 \delta^a_i + \partial_i \psi^a ) \;,
\end{aligned}
\end{equation}   
where $ \psi^a $ is the fluctuation of $ \varphi^a $ on top of background (\ref{Sol_Ill_Ihc}). The only field undergoing SSB is $ \varphi^a $, while $ V^i_a $ and $ \theta $ are spectators and, hence, represent matter fields in the low energy phase. An important thing to note is that $ V^i_a $ is charged under the action of $ SO(d)_{int} $, while a low energy observer will introduce fields as linear representations of $ SO(d)_V $, since only the latter group is unbroken. Hence, one should transform Lagrangian (\ref{Eff_Lagr_S2}) further, so that the matter fields will be charged under the action of $ SO(d)_V $ only.

%%%%%%%%%%%%%%%%%%%%%%%%%%%%%%%%%%%%
\subsubsection{Employing coset space technique}
%%%%%%%%%%%%%%%%%%%%%%%%%%%%%%%%%%%%
\label{sec:2-2-2}

Before finding the required field redefinition, let us discuss the question of which coset space should be used to reconstruct the effective Lagrangian. For this purpose, we apply the formalism of reducing matrix \cite{Salam:1969rq,Weinberg:1996kr}, also known as the polar decomposition, to theory (\ref{Ill_Ihc}) with vacuum expectation value (VEV)  (\ref{Sol_Ill_Ihc}). As we will show, the answer is not the expected one, given by Eq. (\ref{ParamISO}).

The idea of polar decomposition is to separate NGF and other fields:
\begin{equation} \label{Polar_Decomp}
\chi(x) = \gamma(x) \tilde{\chi}(x) \;,
\end{equation}
where $ \chi $ are the fields of a theory under consideration and $ \tilde{\chi}(x) $ is such that it does not include NGF. For theory (\ref{Ill_Ihc}), $ \chi(x) $ and $ \tilde{\chi}(x) $ are introduced as
\begin{equation}
\begin{gathered}
\chi(x) = ( \varphi^1,~ ...~, \varphi^d, \, V^1_1,~ ...~, ~V^d_d, \, \theta )^T \,, 
\\ \tilde{\chi}(x) = ( \tilde{\varphi}^1,~ ...~, \tilde{\varphi}^d,\, \tilde{V}^1_1,~ ...~, \tilde{V}^d_d,\, \tilde{\theta} )^T \;,
\end{gathered}
\end{equation}
Since NGF are DoF associated with the action of the broken generators on the vacuum, the condition for $ \tilde{\chi}(x) $ not to include NGF reads as follows,
\begin{equation} \label{Cond_No_NGF}
\tilde{\chi}^T(x) ( \hat{Z}_a \chi(x) ) = 0 \;,
\end{equation}
where $ Z_a $ are broken generators and $ \hat{Z}_a $ is their representation appropriate for $ \chi(x) $. Taking $ Z_a $ to be the broken internal translations, we get
\begin{equation}
\tilde{\varphi}^a = 0 ~~ \text{for all}~ a \;. 
\end{equation}
Further, taking $ Z_a $ to be $ \bar{M}_{ab} $ yields no additional restriction, since $ V^i_a = 0 $ on the background solution. Thus, we have
\begin{equation}
\tilde{\chi}(x) = (0,~ ... ~, 0,\, V^1_1,~ ... ~, V^d_d, \, \theta) \;,
\end{equation} 
where we have taken into account that decomposition (\ref{Polar_Decomp}) should preserve the number of DoF. Then, knowing the explicit form of $ \chi(x) $ and $ \tilde{\chi}(x) $, from Eq. (\ref{Polar_Decomp}) we find
\begin{equation} \label{Gamma_Explicit}
\gamma(x) = e^{i\bar{P}_a \psi^a(x)} \;.
\end{equation}
Now we substitute (\ref{Polar_Decomp}) back into Lagrangian (\ref{Ill_Ihc}). Remembering further that Lagrangian (\ref{Ill_Ihc}) can be written in terms of the wedge products,\footnote{To obtain the box operator, one would also need to use the $ \mathcal{D} $ operator \cite{Ogievetsky1974,Weinberg:1996kr}. Its construction is a standard part of CST.} and by making use of transitivity, we obtain that the only NGF that is present in the theory is $ \psi^a $, and that it will appear in the effective Lagrangian via the combination
\begin{equation}
e^{-i\bar{P}_a \psi^a(x)} e^{-iP_{Vi} x^i} d e^{iP_{Vi} x^i} e^{i\bar{P}_a \psi^a(x)} \;.
\end{equation}
Hence, to reproduce Lagrangian (\ref{Sol_Ill_Ihc}) one should consider the coset space
\begin{equation} \label{Coset_NoRed}
g_H = e^{iP_i x^i} e^{i\bar{P}_a \psi^a} \;.
\end{equation}
Let us show that such coset space does allow to reproduce effective Lagrangian (\ref{Eff_Lagr_S2}). From (\ref{Coset_NoRed})
one can readily read out the Maurer--Cartan forms,
\begin{equation}
\omega_P^i = dx^i \, , ~~~~ \omega_{\bar{P}}^a = d\psi^a \,, ~~~~ \omega_M^{ab} = \omega_L^{ij} = 0\; .
\end{equation}
The covariant derivative of $ \psi^a $ is then $ D_i \psi^a = \partial_i \psi^a $. Taking a covariant derivative $ \mathcal{D}_j $ of $ D_i \psi^a $ as if it was a matter field  \cite{Ogievetsky1974,Weinberg:1996kr}, one gets $ \mathcal{D}_j D_i \psi^a = \partial_j \partial_i \psi^a $. Then, the part of effective Lagrangian (\ref{Eff_Lagr_S2}) containing $ \psi^a $ is reproduced as
\begin{equation}\label{EffLagr3}
 -\frac{1}{2} ( \mathcal{D}^i D_i \psi^a)^2 + \lambda \theta V^i_j D_i \psi^j  \; .
\end{equation}
Thus, we have reproduced effective Lagrangian (\ref{Eff_Lagr_S2}) within coset space (\ref{Coset_NoRed}).

The considered example admits a straightforward generalization. Suppose one is given fields of a theory $ \chi(x) $ and their VEV. Let $ Z_a $ be a full set of broken generators, $ B_\alpha \in Z_a $ be a subset of $ Z_a $ consisting of all generators acting nontrivially on the VEV at the origin, and let $ S_n $ supplement $ B_\alpha $ to the full set of generators of $ Z_a $. Then, note that a generator $ S \in S_n $ can be broken if and only if there exists $ B \in B_{\alpha} $ such that 
\begin{equation} \label{Red_Gen}
[\tilde{P}_\mu, S] \ni B \;,
\end{equation}
where $ \tilde{P}_\mu $ are translational generators in the broken phase of the theory.\footnote{As we saw on examples of this section, $ \tilde{P}_\mu $ may differ from $ P_\mu $, the translational generators in the unbroken phase.} Indeed, the action of $ S $ at a point $ x^\mu $ is related to its action at the origin by the formula
\begin{equation} \label{STGen}
S(x) = e^{-i\tilde{P}_\mu x^\mu} S(0) e^{i\tilde{P}_\mu x^\mu} \; .
\end{equation}
Since the action of $ S(0) $ on the vacuum is trivial, $ S $ is broken if and only if (\ref{Red_Gen}) holds. Thus, the breakdown of $ S_n $ is always the consequence of the breakdown of $ B_\alpha $. In what follows, we will call generators $ S_n $ partially broken to distinguish them from $ B_\alpha $. Further, consider the analog of Eq. (\ref{Polar_Decomp}) for determining $ \gamma $. Since $ B_\alpha $ are independent, Eq. (\ref{Cond_No_NGF}) with $ Z_a $ taken to be $ B_\alpha $ are independent as well. On the other hand, since the action of $ S_n $ on the vacuum reduces to that of $ B_\alpha $, substituting them into Eq. (\ref{Cond_No_NGF}) does not yield new constraints on $ \tilde{\chi}(x) $. Hence, to exclude the NGF from $ \chi(x) $, it is enough to choose $ \gamma $ in the form
\begin{equation}
\gamma = e^{iB_\alpha \xi^\alpha} \;,
\end{equation}
where $ \xi^\alpha $ are the NGF for $ B_\alpha $. Then, by repeating further steps, one concludes that only $ B_\alpha $ should be included into the coset space, and, hence, only the NGF $ \xi^a $ will be present in the effective theory.

In Appendix \ref{sec:a} we justify the prescription above in the language of induced representations, by employing the connection between them and CST. Therein we also generalize our result to the case when vacuum solution is of the soliton type. Finally, note that from our results it follows that when dealing with the redundant fields, the inverse Higgs phenomenon cannot be considered as a real physical effect, nor as a gauge fixing condition.

%%%%%%%%%%%%%%%%%%%%%%%%%%%%%%%%%%%%
\subsubsection{Interpreting inverse Higgs phenomenon}
%%%%%%%%%%%%%%%%%%%%%%%%%%%%%%%%%%%%
\label{sec:2-2-3}

From the discussion above we conclude that with the proper usage of the CST one can avoid introducing redundant NGF at any step of the construction of an effective theory. However, the effective Lagrangians obtained in this way do not provide the required from the low energy perspective parametrization of DoF: fields are charged under the action of partially broken generators, which is not the way a low energy observer will introduce them. Hence, one should find a way to ``uncharge'' matter fields under the action of partially broken generators. For the theory under consideration, this implies that one should search for a field redefinition
\begin{equation} \label{Uncharge_Fields}
V^i_a ~ \rightarrow ~ \Omega^b_a \tilde{V}^i_b \;,
\end{equation}
where $ \Omega^b_a $ belongs to $ SO(d)_{int} $ and is a function of $ \psi^a $, the NGF at hand. If it is possible to compose $ \Omega^b_a $ from $ \psi^a $, this will allow us to express the transformation of $ V^i_a $ under the action of $ SO(d)_{int} $ through the transformation of $ \Omega^b_a $. This will also allow to uncharge $ V^i_a $ under the action of $ SO(d)_{ST} $, since any such transformation can be completed to a composition of the diagonal and internal transformations. Hence, the question is whether a suitable matrix $ \Omega^b_a $ exists. 

Let us show that the answer to the question above is positive and the procedure of uncharging $ V^i_a $, in fact, corresponds to employing inverse Higgs phenomenon. To this end, note that if the fields $ \omega^{ab} $ were true NGF, $ \Omega^a_b $ could be taken as $ \Omega(\omega)^a_b $. One should therefore find a combination of $ \psi^a $ that has the same transformation properties as $ \omega^{ab} $. This can be achieved by the means of coset (\ref{ParamISO}) in which $ \omega^{ab} $ are considered as auxiliary fields, and the desired combination of $ \psi^a $ is provided by the IHC. Thus, one can use the latter to make the field redefinition (\ref{Uncharge_Fields}).

Note that the suitable expression for the matrix $ \Omega^a_b $ can be found within any coset space, including $ \omega^{ab} $, not necessarily the one given in Eq. (\ref{ParamISO}). The latter choice, however, provides the most convenient expression. Indeed, depending on the parametrization, $ \Omega^a_b $ can in general include coordinates and fields other than $ \psi^a $ \cite{Klein:2017npd}. As our analysis shows, the Lagrangians obtained via different parametrizations of the coset are equivalent, and one can switch between them by suitable fields redefinitions.

It is instructive to point out the difference in the parametrizations of DoF in the theories of this and the previous sections. As we showed, in the current example employing inverse Higgs phenomenon amounts to redefining the DoF (\ref{Uncharge_Fields}). In the example of Sec. \ref{sec:2-1} the analogous redefinition for the field with its ``inverse Higgs part'' extracted takes the form
\begin{equation}
V^i_a \rightarrow \Omega^c_a (\psi) \Omega^b_c(\tilde{\omega}^{kl}) \delta^i_b M \;,
\end{equation}
where $ \tilde{\omega}^{kl} = \tilde{A}^{kl}/M $. Hence, the difference is that the first theory requires the presence of $ \tilde{\omega}^{ab} $ in order to describe the full set of fluctuations of the vacuum, while the second theory does not.

%%%%%%%%%%%%%%%%%%%%%%%%%%%%%%%%%%%%
\section{Important consequences} 
%%%%%%%%%%%%%%%%%%%%%%%%%%%%%%%%%%%%
\label{sec:3}

%%%%%%%%%%%%%%%%%%%%%%%%%%%%%%%%%%%%
\subsection{Nambu-Goldstone theorem and structure of effective theories}
%%%%%%%%%%%%%%%%%%%%%%%%%%%%%%%%%%%%
\label{sec:3-1}

Let us summarize briefly the results of the previous section. First, we showed that all broken generators acting nontrivially on a vacuum at the origin (and only they) give rise to independent NGF\footnote{This rule applies when the dimensions of the full and effective theories are equal, i.e., when the number of unbroken translation generators in these phases are equal. The generalization of this result to the situations when this is not the case is given in Appendix \ref{sec:a}.}. This constitutes the Nambu--Goldstone theorem for SSB of spacetime symmetries. Second, for each partially broken generator one should introduce an auxiliary field that is excluded from the content of the theory with the help of IHC. Note that Eq. (\ref{Red_Gen}) guarantees that this step can always be performed. Different parametrizations of the coset space result in different ways to eliminate redundant fields in favor of independent NGF, but all of them are physically equivalent. 

To find the number of mNGF, note that in the course of applying the polar decomposition all $1$--forms containing NGF without differentials can appear only from commuting out the term
\begin{equation} \label{CommutOutTransl}
e^{-iZ_a\xi^a} (e^{-i\tilde{P}_\mu x^\mu} d e^{i\tilde{P}_\mu x^\mu}) e^{iZ_a\xi^a} ~ \in ~ g_H^{-1} d g_H  \;.
\end{equation}
As it was proven in \cite{Nicolis:2013sga}, the derivative coupling between various NGF does not change the total number of gapped and gapless states. Consequently, there are as many gapped fields as many NGF enter Eq. (\ref{CommutOutTransl}) without differentials, which reduces the problem to the explicit calculation of (\ref{CommutOutTransl}). The only possible exception to this rule are the NGF entering the $1$--form for the translations without differential --- depending on the group under consideration, they may, in fact, disappear from the effective metric, which would yield them massless. Hence, the question of masslessness of such modes should be addressed separately. 

Due to the possible presence of mNGF, the structure of effective theories arising from SSB of spacetime symmetries is, in general, qualitatively different from that corresponding to SSB of internal symmetries. In both cases, a low energy theory contains the strong coupling scale at which it must be UV completed. But when spacetime symmetries are involved in SSB, the scales associated with mNGF will appear as well. At energies much below these scales, the corresponding mNGF can be integrated out. Generally, the relation between the strong coupling scale and the scales of mNGF can be arbitrary. For example, in the theory of Sec. \ref{sec:2-1}, the strong coupling scale is $ M $, while the mass of the vector field, $ \varkappa $, can, in fact, be of the order of or even larger than $M$. If $ \varkappa \gtrsim M $, the dynamics of the vector field can be neglected all the way up to the UV cutoff of the effective theory. In any case, in a low energy limit mNGF become inessential \cite{Brauner:2014aha,Klein:2017npd}. However, their presence is important from the perspective of a UV completion which cannot be successfully made without adding mNGF at some stage \cite{Endlich:2013vfa}.

Further, following \cite{Brauner:2014aha}, we note that mNGF can always be redefined to transform linearly under the action of the full symmetry group. Indeed, suppose we have an mNGF $ A $, which enters some homogeneously transforming Maurer--Cartan form $ \Omega_A $ without differential (we have dropped indices for simplicity). Then, one can make a change of variables
\begin{equation}
A' = \Omega_A \;.
\end{equation}
This represents a valid change of variables because both sides of the equation contain fields without differentials. Next, since Maurer--Cartan form $ \Omega_A $ transforms homogeneously, so will $ A' $. In this parametrization, $ A' $ transforms like ordinary matter fields and represents massive nonradial modes observed in \cite{Endlich:2013vfa}. The formulated above analog of the Nambu--Goldstone theorem for SSB of spacetime symmetries establishes the criterion when such modes will be present in the effective theory, which was missing in \cite{Brauner:2014aha}. 

We would also like to note that from our analysis it follows that if some subgroup of a symmetry group acts trivially on fields at the origin, the generators of this subgroup never give rise to NGF \cite{Hidaka:2014fra}. As an instructive example, consider the conformal group and let $ K_n $ be the generators of special conformal transformations. Then, since $ \hat{K}_n \Phi = 0 $ for a quasiprimary $ \Phi $, they do not describe independent fluctuations of the vacuum. Hence, for example, the $\text{Conf}\rightarrow ISO(1,d)$ SSB pattern can give rise only to a single NGF corresponding to the broken dilations. This implies that the NGF for special conformal transformations are always auxiliary and must be excluded by employing the inverse Higgs phenomenon, which agrees with the result of \cite{Kharuk:2017jwe,Kharuk:2017gcx}. Also, the outlined above consequence may be of interest in the context of polynomial symmetries \cite{Nicolis:2008in,Griffin:2014bta}. 

%%%%%%%%%%%%%%%%%%%%%%%%%%%%%%%%%%%%
\subsection{Comparison with the literature}
%%%%%%%%%%%%%%%%%%%%%%%%%%%%%%%%%%%%
\label{sec:3-2}

We would like to start this section by making contact with \cite{Low:2001bw,Nicolis:2013sga}. As it was shown in these works, redundancies among NGF appear when some of the broken generators do not produce independent fluctuations of a background configuration. Let us illustrate this phenomenon using theories (\ref{Largr_mNGF}) and (\ref{Ill_Ihc}) as examples. Denote by $\Phi$ the collection of fields forming the background, and consider the following equation on $\Delta \psi^a$, $\Delta\omega^a_b \,$,
\begin{equation} \label{RedSearch}
\delta \Phi \equiv (\Delta \psi^a \bar{P}_a + \frac{1}{2}\Delta\omega^a_b \bar{M}^b_a ) \Phi = 0 \; .
\end{equation}  
If nontrivial solutions of this equation exist, then the field configurations $ (\psi^a, \omega^a_b) $ and $(\psi^a +\Delta\psi^a, \omega^a_b + \Delta\omega^a_b)$ describe the same fluctuation of the background, and, hence, the set of variables $ (\psi^a, \omega^a_b) $ is redundant. It is easy to see that for background (\ref{VEV_mNGF}) there are no nontrivial solutions to Eq. (\ref{Sol_Ill_Ihc}). On the other hand, with zero value of the vector field, Eq. (\ref{Sol_Ill_Ihc}), the solution is of the form
\begin{equation}\label{SimTransf}
\Delta \psi^a(x) = x^b \alpha^a_b(x) \, , ~~~~ \Delta \omega^a_b(x) = \alpha^a_b(x) \; ,
\end{equation}
where $ \alpha^a_b(x) $ is an arbitrary antisymmetric tensor field. The existence of such solutions for theory (\ref{Ill_Ihc}) reflects the fact that $ \bar{M}^a_b $ are partially broken --- they annihilate the vacuum at the origin but commute with the unbroken translations to the broken $ \bar{P}_a $. This is the reason why Eq. (\ref{RedSearch}) has nontrivial solutions. As our analysis shows, the NGF $ \omega^{ab} $ are auxiliary fields and should not be interpreted as physical DoF. Note that one can come to the same conclusion by noticing that it is possible to nullify $ \omega^{ab} $ in the whole spacetime by choosing $ \alpha^{ab} $ properly. On the other hand, an attempt to nullify $\psi^a$ fails since the corresponding function $\alpha^a_b$ will be singular at the origin of coordinates.

Next, we would like to note that our construction is in a full agreement with the results of \cite{Watanabe:2013uya}. There, it was noticed that if there are functional relations between Noether currents associated with broken symmetries, then the corresponding NGF are redundant and should not be introduced as independent fields. In our approach, these redundant NGF are identified as auxiliary fields from the very beginning. Although the reasoning leading to this equivalence is very similar to the previous one, let us provide it for illustrative purposes. Consider a scalar field theory with a coordinate--dependent (say, $z$--dependent) VEV. Then, since the Lorentz generators act trivially on the scalar field at the origin, they are only partially broken and, hence, the corresponding NGF are auxiliary. To arrive at the same result by using the method of \cite{Watanabe:2013uya}, note that the action of the Lorentz group in the whole spacetime is given by formula (\ref{STGen}). Since $ M_{z\mu} $ and $ P_{\nu} $ commute to the broken translation $ P_z $, the Lorentz generators act nontrivially at a general spacetime point. Combined with the trivial action at the origin, this results in the following functional dependence of the energy--momentum and the angular momentum tensors,
\begin{equation}
M^\lambda_{z\nu} = z T^\lambda_\nu - x_\nu T^\lambda_z \;.
\end{equation}
Hence, according to \cite{Watanabe:2013uya}, NGF for the Lorentz group are redundant and one should not introduce them as independent fields. The generalization of this example to general case is straightforward.

Finally, we would like to make contact with \cite{Endlich:2013vfa}. From the discussion above we see that what was argued to be a new strong coupling scale in this work is nothing but the scale at which the effect of mNGF on the low energy physics cannot be neglected anymore. Namely, since the Lorentz and internal rotations were broken down to the diagonal subgroup, and since the action of the internal group at the origin is nontrivial, the corresponding NGF are physical. Hence, the theory studied in \cite{Endlich:2013vfa} must include mNGF. In particular, since these fields can be defined to transform linearly under the action of the full symmetry group, they were not recognized as the fields needed to restore the broken symmetries.

%%%%%%%%%%%%%%%%%%%%%%%%%%%%%%%%%%%%
\subsection{Inverse Higgs phenomenon in massive gravity}
%%%%%%%%%%%%%%%%%%%%%%%%%%%%%%%%%%%%
\label{sec:3-3}

The obtained results allow us to reveal versions of massive gravity that have not been studied so far. Namely, to restore diffeomorphism and local Lorentz invariance, one usually introduces 10 Stukelberg fields \cite{Ondo:2013wka,Gabadadze:2013ria}: 4 scalars $ \varphi^a $, restoring diffeomorphism invariance, and 6 antisymmetric spin--2 fields $ \Lambda_{\mu\nu} $ restoring local Lorentz transformations.\footnote{In \cite{Goon:2014paa} the symmetries are restored by introducing 16 fields. We do not consider this case here, though our discussion applies to it as well.} However, the results of Sec. \ref{sec:2} give us a hint that the field $ \Lambda_{\mu\nu} $ may not be independent, and, hence, the full invariance can be restored by only 4 fields, as it was suggested in \cite{Dubovsky:2004sg}. To show that such theories are possible, consider a typical term appearing in massive gravity \cite{deRham:2010ik,deRham:2010kj,deRham:2011rn},
\begin{equation} \label{dRGTTerm}
\mathcal{L}_{dRGT} = \epsilon_{abcd}\, 1^a \wedge e^b \wedge e^c \wedge e^d \;,
\end{equation}
where $ 1^a = \delta^a_\mu dx^\mu $ and $ e^a $ are the tetrads. The 1--form $ 1^a $ plays a central role in this construction and is responsible for the breakdown of diffeomorphism and local Lorentz invariance. Note, however, that Lagranian (\ref{dRGTTerm}) is invariant under the diagonal subgroup of the Lorentz groups acting on tetrad and spacetime indices. Hence, it corresponds to the following ``SSB pattern'' \cite{Goon:2014paa},
\begin{equation} \label{PatternMG}
SO_{gc} \times SO_{loc} \rightarrow ( SO_{gc} \times SO_{loc} )_{diag} \;,
\end{equation}
where $ SO_{gc} $ and $ SO_{loc} $ act on spacetime and tetrad indices respectively. This situation is analogous to the one we studied in the examples of Sec. \ref{sec:2}, where it was possible to realize the similar pattern with the different amount of fields. In particular, if we have at hand a field $ \varphi^a $ transforming as a vector in the broken phase, then one can introduce an auxiliary field
\begin{equation} \label{CompensLLTviaScalars}
\omega_{\mu\nu} = \partial_{[\mu} \varphi_{\nu]} \,, ~~~ \Lambda_{\mu\nu} = e^{\omega_{\mu\nu}} \;,
\end{equation} 
which allows to realize all of the symmetries by only 4 Stuckelberg fields. This would be the case when $ \Lambda_{\mu\nu} $ is not associated with physical DoF. In the context of decoupling limit and other aspects of massive gravity, this possibility has not yet been studied in the literature. In particular, the analysis of \cite{Ondo:2013wka,Gabadadze:2013ria} does not cover this case since it is not valid to vary the action with respect to $ \Lambda_{\mu\nu} $ when the latter is given by (\ref{CompensLLTviaScalars}). We leave any more detailed consideration of such theories for elsewhere.

Finally, we would like to mention the similarity between the model of Sec. \ref{sec:2-1} and that of \cite{Blas:2014ira}. The latter work is devoted to the UV completion of Lorentz--violating massive gravity, and its matter sector has the same structure as the example of Sec. \ref{sec:2-1}. More precisely, the scalar and bi--fundamental fields of \cite{Blas:2014ira} acquire a time--dependent VEV very similar to Eq. (\ref{VEV_mNGF}), which allows to give a mass to the graviton in the IR phase. We conclude that theories admitting mNGF can play an important role in studies of possible UV completions and IR modifications of general relativity. 

%%%%%%%%%%%%%%%%%%%%%%%%%%%%%%%%%%%%
\section{Conclusion}
%%%%%%%%%%%%%%%%%%%%%%%%%%%%%%%%%%%%
\label{sec:4}

In this paper, we established the Nambu--Goldstone theorem for SSB of spacetime symmetries. Namely, we showed that the careful use of the polar decomposition uniquely fixes the NG sector of a theory. All broken generators acting nontrivially on a vacuum at the origin give rise to independent NGF, while the remaining fields are auxiliary. Massive nonradial modes, which one may have to introduce to UV complete a theory resulting from SSB of spacetime symmetries, are nothing but mNGF. They can be made to transform linearly under the action of the full symmetry group.

We also clarified the physical meaning of the inverse Higgs phenomenon. Contrary to often seen interpretation, its aim is not to reduce the number of DoF in the effective theory. Instead, it is used to find all NGF that transform nonlinearly under the action of the broken generators. The other NGF may or may not be present in the theory, depending whether they are needed or not to complete the set of possible fluctuations of the vacuum. In particular, when the CST is used for obtaining Lagrangians with gauge invariance \cite{Ivanov:1976zq,Goon:2014ika,Ivanov:1981wn,Goon:2014paa}, following inverse Higgs phenomenon accounts for not using some of the modes. The obtained insight into the meaning of the inverse Higgs phenomenon and possible presence of mNGF can be relevant in massive gravity \cite{Goon:2014paa} and in a so--called self--gravitating medium \cite{Ballesteros:2016gwc}.

\section*{Acknowledgments}
The authors thank E. Ivanov, R. Penco and S. Sibiryakov for useful discussion. The work of I.K. was supported by the Grant 14-22-00161 of the Russian Science Foundation.

\appendix

%%%%%%%%%%%%%%%%%%%%%%%%%%%%%%%%%%%%
\section{Coset construction revisited}
%%%%%%%%%%%%%%%%%%%%%%%%%%%%%%%%%%%%
\label{sec:a}

In this Appendix we provide a mathematical justification of the rule formulated in Sec. \ref{sec:2-2} for determining which generators must be present in the coset space. As we demonstrate, the rule follows directly from the method of induced representations. Also, here we provide the generalization of our counting rule of NG fields for effective theories defined in the reduced number of dimensions.

%%%%%%%%%%%%%%%%%%%%%%%%%%%%%%%%%%%%
\subsection{Induced representations}
%%%%%%%%%%%%%%%%%%%%%%%%%%%%%%%%%%%%
\label{sec:a-1}

We start with a brief outline of the method of induced representations \cite{Mackey:1969vt,hermann1966lie,ortin2004gravity}, which allows us to fix the notations and to remind the underlying structure of the construction. Let $ G $ be a symmetry group,\footnote{Below we assume that the action of $ G $ is global and that $ G $ does not include discrete elements.} $ \A $ its chosen homogeneous space, and $ H $ a stability group of some point $ \vec{0} $ in $ \A $. Since $ \mathcal{A} $ is the homogeneous space of $ G $, there is a one--to--one correspondence,
\begin{equation} \label{IsomHomSpace}
\A = G/H\;. 
\end{equation}
Denote by $ V_i $ the generators of $ H $ and by $ P_\mu $ the rest of the generators of $G$. Then, Eq. (\ref{IsomHomSpace}) establishes the isomorphism between $ \mathcal{A} $ and the orbit of $ \vec{0} $ under the action of an element $ g_H $ of the coset space $ G/H$,
\begin{equation} \label{CosetInduced}
g_H = e^{iP_\mu x^\mu}\;.
\end{equation}
Within this isomorphism, an arbitrary element of $G/H$ is identified with the point of $ \mathcal{A} $ obtained by acting by the former on $ \vec{0} $. As $ g_H $ is uniquely characterized by $ x^\mu $, it is natural to refer to $ P_\mu $ as generators of translations and to $ x^\mu $ as coordinates on $ \A $. Consider further the left action of $ G $ on $ G/H $, which for arbitrary $ g \in G $ can be written as\footnote{Considering the right action of $G$ on $G/H$ would lead to an equivalent representation of $G$.}
\begin{equation} \label{ArbAction}
g \cdot g_H = g'_H (g,g_H) \cdot h(g,g_H)\;,
\end{equation}  
with $ h(g,g_H) \in H $. This naturally defines the transformation rule of $ g_H $ under the action of $ G $ to be
\begin{equation} \label{IndActCoord}
g_H \rightarrow g \cdot g_H \cdot h^{-1}(g, g_H):  ~~~  x^\mu \rightarrow x'^\mu (g, x^\mu)\;,
\end{equation}
which can be thought of as a change of coordinates. 

Given the space $\A$ and the action of $G$ on its coordinates, we can introduce fields that are defined on $\A$ and form a representation of $ G $. This is done by the method of induced representations, which goes as follows. First, consider a vector space $\mathcal{V}$ on which $H$ acts by some linear representation $T=T(h)$,
\begin{equation} \label{NoInd}
T(h): ~~ \mathcal{V} \rightarrow \mathcal{V}: ~~ \forall \psi \in \mathcal{V} \rightarrow T(h) \psi \;. 
\end{equation}
As the next step, $\psi$'s are promoted to functions with the domain $\A$, taking values in $\mathcal{V}$,
\begin{equation} \label{IndSpace}
\psi \rightarrow \psi(x) \;,
\end{equation}
where we used the fact that each representative of $ G/H $ is uniquely determined by the values of $ x^\mu $. Finally, one defines the action of $ G $ on this space of functions to be
\begin{equation} \label{IndActonMatter}
T(g) \psi(x) = T \left( h^{-1} (g^{-1},g_H) \right)  \psi( x'(g^{-1}, x) )\;,
\end{equation}
where $ h $ is defined from Eq. (\ref{ArbAction}) for $ g_H $ taken at the point $ x^\mu $. The obtained representation of $G$, acting on $ x^\mu $ and $\psi(x)$ via eqs. (\ref{IndActCoord}) and (\ref{IndActonMatter}) accordingly, is called the induced representation. In particular, it can be verified that this is indeed a (nonlinear) representation of $ G $ on the space of $\mathcal{V} $--valued functions on $ \mathcal{A} $.  

An illustrative example of the application of the method of induced representations is the construction of representations of the Poincare group from those of the Lorentz subgroup. In this case, one has $ G= ISO(1,d),~ \A =\M^{1,d}, ~H=SO(1,d) $ and $ P_m $ are the usual translation generators. To obtain a representation of the Poincare group, one first introduces a representation of the Lorentz group, which is characterized by spin. Then, the elements of this representation are promoted to dynamical fields by making them functions of $x^\mu$, thus forming the space of the representation of the Poincare group. Finally, one defines the action of the latter on the coordinates and fields according to Eqs. (\ref{IndActCoord}) and (\ref{IndActonMatter}), which results in the usual well--known expressions. As another example, the same procedure can be applied to the construction of representations of the AdS group, which corresponds to inducing representations of $ SO(1,d) $ to those of $ SO(2,d) $. In this case we have $ G=SO(2,d), ~\A=AdS^{1,d}$ and $ P_n = M_{-1,n} $ \cite{ortin2004gravity}.

The outcome of the discussion above is that the fields are introduced via the two--step construction. First, one introduces a representation of $ H $ on a vector space $\mathcal{V}$, without appealing to $ \A $ in any way. And only when this representation is induced to that of $ G $, do the vectors $\psi$ of $\mathcal{V}$ are replaced by the fields with the domain $ \A $. Note that $\psi$'s can be regarded as the fields defined at the single point $ \vec{0}$ of  $\A $, since at this point the induced representation, Eq. (\ref{IndActonMatter}), reduces to the initial one, Eq. (\ref{NoInd}). Speaking loosely, the induction of the representation amounts to extending the domain of $\psi$ from $ \vec{0} $ to the entire $\A$ in a consistent way.

%%%%%%%%%%%%%%%%%%%%%%%%%%%%%%%%%%%%
\subsection{Induced representations and SSB}
%%%%%%%%%%%%%%%%%%%%%%%%%%%%%%%%%%%%
\label{sec:a-2}

Let us now apply this construction in the case when some of the symmetries are spontaneously broken. That is, consider a theory defined on $ \mathcal{A} $, which develops some nonzero VEV. Then, from the geometric perspective this implies that the effective theory describing the fluctuations on top of this background is defined on $ \A $ accompanied with the VEV of the fields at each point. We will call this space $\tilde{\A}$. For example, if some scalar $ \varphi $ develops constant VEV $ \varphi_0\,, ~ \tilde{\A} $ is a set of points $ (x^\mu, \varphi_0) $. In the case of SSB of spacetime symmetries, the situation gets complicated by the fact that fields are allowed to have coordinate--dependent VEV. Let us illustrate this subtlety using the theory of Sec. \ref{sec:2-1} as an example. In this case the space $ \tilde{\A} $ consists of the points $ (x^\nu, ~\mu^2 x^\nu ,~ M\delta^\nu_a) $. Clearly, the action of $ P_\mu $ on this space is not transitive, and, hence, one should search for new ``effective'' translational generators $ \tilde{P}_\mu $ that would act transitively on $ \tilde{\A} $. For the case under consideration, they are formed by the unbroken combinations of the internal and spacetime translations, which we used in coset space (\ref{ParamISO}). Note, however, that, in general, the generators $ \tilde{P}_\mu $ acting transitively on $ \tilde{\A} $ may not exist. For example, this situation takes place for the scalar domain wall. Postponing the discussion of this possibility to section \ref{sec:a-3}, here we assume that this is not the case.

To proceed further, let us fix $ \tilde{\A} $ to be a set of points $ (x^\mu, \varphi_\alpha(x)) $ for some $ \varphi_\alpha(x) $ ($ \alpha $ can stand for spacetime or internal indices), $ H_0 $ to be the stability group of $ \tilde{\A} $ at $ \vec{0} $, and $ Z_a $ to be the set of generators supplementing $ \tilde{P}_\mu $ and the generators of $ H_0 $ to the full set of generators of $ G $. Then, note that the range of the fluctuations of the background at $ \vec{0} $ is formed by the points $ (\vec{0}, \psi) $ with all possible values of $ \psi $. To reflect this fact, we introduce the quotient space of $G$ by $ (H_0 \times \A ) $,
\begin{equation} \label{StabBroken}
g_{H_0} = e^{iZ_a \xi^a} \;.
\end{equation}
This step is very similar to introducing the coordinates on $ \A $ via coset (\ref{CosetInduced}), except for the fact that the action of coset (\ref{StabBroken}) on $ \tilde{\A} $ at $ \vec{0} $ establishes the isomorphism between $ \xi^a $ and $ \psi $. Importantly, (\ref{StabBroken}) yields all of the NGF (which at this stage are vectors) that must be present in the theory to realize $G$ nonlinearly. The action of $ h \in H $ on $ \xi^a $ is realized as the left action of $ H $ on coset (\ref{StabBroken}) , 
\begin{equation} \label{IndNGAct}
g_{H_0} \rightarrow h \cdot g_{H_0} \cdot h_0^{-1}(h, g_{H_0}):  ~~~  \xi^a \rightarrow \xi'^a (h, g_{H_0})\;,
\end{equation}
where $ h_0 \in H_0 $ is such that $ h \cdot g_{H_0} = g'_{H_0} \cdot h_0 $. In particular, the action of all $ h_0 \in H_0 $ on $ \xi^a $ is linear, as it follows from 
\begin{equation}
h_0 \cdot g_{H_0} = (h_0 \cdot g_{H_0} \cdot h^{-1}_0) \cdot h_0\;.
\end{equation}
After obtaining this representation of $ H $, one should induce it to that of $ G $. This is done by introducing the exponentials of the effective translations to coset (\ref{StabBroken}) and promoting $ \xi^a $ to functions of $ x^\mu $,
\begin{equation} \label{SSBCoset}
g_H = e^{i\tilde{P}_\mu x^\mu} e^{iZ_a \xi^a(x)} \; .
\end{equation}
Once $ \xi^a $ become fields in this way, they can be identified with the NGF corresponding to the broken generators. 

As we see, the accurate use of the method of induced representations shows that one should include into the coset only the generators acting nontrivially on the vacuum at the origin. This provides an independent justification of our approach to the construction of effective Lagrangians resulting from SSB of spacetime symmetries within the CST.

%%%%%%%%%%%%%%%%%%%%%%%%%%%%%
\subsection{Embedding case}
%%%%%%%%%%%%%%%%%%%%%%%%%%%%%
\label{sec:a-3}

Consider now the case when a set of generators acting transitively on $ \mathcal{\tilde{A}} $ does not exist. To understand how the CST should be applied in this case, let us consider a theory admitting scalar domain wall background. The simplest theory admitting DW configuration reads
\begin{equation} \label{DWLagr}
\mathcal{S} = \int d^dxdz \left( \frac{1}{2}\partial_m \varphi \partial^m \varphi - \frac{\lambda}{4} \left( \varphi^2 - v^2 \right)^2 \right) \; ,
\end{equation}
where $\varphi$ is a real scalar field and $\lambda,v>0$. By choosing suitable coordinates, a general DW solution can be brought to the form
\begin{equation} \label{ScalDWSolut}
\varphi_z = v \tanh \left( \sqrt{\frac{\lambda}{2}}vz \right) \; .
\end{equation}
Expanding theory (\ref{DWLagr}) around solution (\ref{ScalDWSolut}) at quadratic level, we find one massless and one massive excitations. The first mode appears as a result of the breakdown of the translational invariance along $z$--direction, while the second represents a massive bound state.\footnote{Note that this mode does not correspond to the broken Lorentz transformations. In fact, the number of massive excitations depends on the form of the potential for $\varphi$ and can be cast to zero.} The effective Lagrangian for perturbations is of the Nambu--Goto type,
\begin{equation} \label{DirectEffAction}
\begin{gathered}
\mathcal{L}_{\psi} = \int dz (\partial_z\varphi_z)^2 \sqrt{|h|}\,, \\ 
h = \text{det}\, h_{ij}\,, ~~~ h_{ij} = \eta_{ij} + \partial_i \psi \partial_j \psi \;,
\end{gathered}
\end{equation}
where $ \psi=\psi(x) $ is the difference between $ \varphi $ and $\varphi_z$ and $ \eta_{ij} $ is the Minkowski metric. 

For this theory, the space $\tilde{\A}$ is given by the set of points $ (x^\mu\,,\, z\,,\, \varphi_z(z) ) $. Clearly, the orbit of none of these points under the action of $ ISO(1,d) $ spans the whole $ \mathcal{\tilde{A}} $. 
However, $ \mathcal{\tilde{A}} $ can be covered by an orbit of the surface
\begin{equation} \label{SurfaceForDW}
\mathcal{B} = \lbrace \, (0\,, z\,, \varphi_z(z) )\,, \; z\in (-\infty, +\infty ) \, \rbrace \; 
\end{equation}
under the action of $ P_\mu $. This suggests the following way of constructing the effective action. First, one applies the CST to obtain a $d$--dimensional effective theory at a given point of $ \mathcal{B} $. By construction, this gives the Lagrangian density of the full $(d+1)$--dimensional effective theory taken at this point. Then, one builds such densities at all of the points of $ \mathcal{B} $ and integrates over $z$, which yields the full effective Lagrangian. 

Let us elaborate on how these effective Lagrangian densities should be obtained. Note first that applying the method of induced representations at each point of $ \mathcal{B} $ does not allow to reproduce the group action along $z$--direction. However, the latter was already defined in the unbroken phase, and one should take it as the definition of the group action in the spontaneously broken phase. Such prescription is free of inconsistencies because the group action, in the CST framework, is uniquely fixed by the group multiplication law. Then, for a given point $z$ of $ \mathcal{B} $, denote by $ \mathcal{B}_z $ and $H_z$ its orbit under the action of $ P_\mu $ and its stability group accordingly. Since the full effective theory is recovered by integrating over $z$, the stability group $ H_{\mathcal{B}} $ of the full theory is the intersection of all $H_z$. Further, the Lagrangian densities are obtained as the embeddings of $ \mathcal{B}_z $ into $ \mathcal{\tilde{A}} $. Finally, by integrating these densities over $z$, one finds the full effective Lagrangian.

For a scalar domain wall background, $ H_{\mathcal{B}} $ is the Lorentz group. Then, since $ P_\mu $ form the effective $d$--dimensional translations, we conclude that the only broken generator is $ P_z $. Hence, to obtain the ingredients for the construction of the effective action, one should consider the SSB pattern
\begin{equation} \label{DWAlongZ}
ISO(1,d) \rightarrow e^{iP_i x^i} \times SO(1,d) \; ,
\end{equation}
with $ P_z $ giving rise to the NGF $ \psi =\psi(x) $, while the rest of the translation generators give rise to the coordinates. Let us show that such prescription indeed allows one to reproduce effective action (\ref{DirectEffAction}). The coset space corresponding to SSB pattern (\ref{DWAlongZ}) reads
\begin{equation}\label{DW_CosetSpace}
g_H = e^{iP_i x^i}e^{iP_z\psi} \;,
\end{equation}
and the Maurer--Cartan forms are easily found to be
\begin{equation} \label{MCFSDW}
\omega^i_P = dx^i \, , ~~~~ \omega^z_P = d\psi \;.
\end{equation}
The effective action of the theory is the action of the fluctuating domain wall embedded into the bulk Miknowski spacetime. Consequently, all invariant quantities used to build the effective theory should be projected from the bulk to the surface of the DW. In particular, the projected invariant volume form reads \cite{Gabadadze:2013ria}
\begin{equation} \label{LagrTetrads}
\begin{gathered}
\text{Vol}. = \epsilon_{i_1 .. i_d} (\lambda_{n_1}^{i_1} dx^{n_1} ) \wedge ... \wedge (\lambda_{n_d}^{i_d} dx^{n_d} ) \,, \\
\lambda^i_j = \delta^i_j\,, ~~ \lambda^i_z = \eta^{ij} \frac{\delta}{\delta dx^j} \;,
\end{gathered}
\end{equation}
where $ \lambda^i_n $ are the projection operators. From now on, we use $ x^n $ and $ y^i $ to denote coordinates in the bulk and on the DW correspondingly, chosen such that $ y^i=x^i $. The tetrads on the DW are defined from the relation
\begin{equation} 
\tilde{e}^i_j dy^j = \lambda^j_n dx^n  \;.
\end{equation}
As it can be verified, this leads to the standard expression for the induced metric,
\begin{equation} \label{MetrInd}
h_{ij} = \eta_{mn}  \frac{\partial x^m}{\partial y^i} \frac{\partial x^n}{\partial y^j} = \eta_{ij} + \partial_i \psi \partial_j \psi \;.
\end{equation}  
Since the invariant volume form (\ref{LagrTetrads}) is the determinant of $ \tilde{e}^i_j $, which, in turn, is a square root of the determinant of the induced metric, at a given $ z $ we have the Lagrangian density
\begin{equation}\label{ScalarDW_EffLagr_CST}
\La_{\psi}^{(z)} = C(z)\sqrt{|h|} \; ,
\end{equation}
where $ C $ is, in general case, a function of $ z $. By integrating these densities over $z$, one finds the full effective Lagrangian,
\begin{equation}
\mathcal{L}_{\psi} = \int dz \: C(z) \sqrt{|h|} \; ,
\end{equation}
Finally, by setting 
\begin{equation}
C(z) =  ( \partial_z \varphi_z )^2
\end{equation}
we recover the correct low energy description of fluctuations above the scalar domain wall by the means of coset space (\ref{DW_CosetSpace}). We would like to note that Lagrangian (\ref{ScalarDW_EffLagr_CST}) is the only one compatible with the Poincare symmetry to the leading order in the covariant derivative of $ \psi $, hence imposing IHC in this case cannot but give the correct answer as well.

The describe procedure above can be generalized as follows. Whenever the generators acting transitively on $\tilde{\A}$ do not exist, one can find the surface $\mathcal{B}$ of minimal dimension $n$ and the generators $ \tilde{P}_\mu $ such that the orbit of $ \mathcal{B} $ under the action of $ \tilde{P}_\mu $ is the whole $ \mathcal{\tilde{A}} $. The action of the group along the $n$ dimensions of $ \mathcal{B} $ should be taken the same as in the unbroken phase. Further, $ \tilde{P}_\mu $ form the effective translations, whose action at a given point $b$ of $ \mathcal{B} $ gives a $ (d-n) $--dimensional surface $ \mathcal{B}_b\, $. The stability group $H_{\mathcal{B}}$ of the effective theory is given by the intersection of the stability groups of all points of $\mathcal{B}$, $ H_{b} $. Then, at each point of $ \mathcal{B} $, one should take a quotient of $G$ by $H_{\mathcal{B}}$ to obtain the broken generators and use CST to build $(d-n)$--dimensional Lagrangian densities. Finally, by integrating them over $ \mathcal{B} $, one recovers the full effective Lagrangian. 

The corresponding Lagrangian densities are obtained as embeddings of $ \mathcal{B}_b $ into $ \mathcal{\tilde{A}} $. Namely, let $ y^\mu $ be the local coordinates on $ \mathcal{B}_b $. Then, the equations $ x^m=x^m(y) $ define the embedding law of $ \mathcal{B}_b $ into $ \mathcal{\tilde{A}} $. This allows to define all invariant objects needed to construct the Lagrangian density on $ \mathcal{B}_b $ as projections of the corresponding objects in $\tilde{\A}$. For example, an invariant volume in $ \mathcal{B}_b $ is given by \cite{Gabadadze:2013ria}
\begin{equation}
\text{Vol}. = \epsilon_{\mu_1 .. \mu_d} \lambda_{n_1}^{\mu_1} ... \lambda_{n_d}^{\mu_d} \omega_P^{n_1} \wedge ... \wedge \omega_P^{n_d} \;,
\end{equation} 
where $ \lambda^\mu_n $ are operators projecting tangent vectors from $ \mathcal{\tilde{\A}}$ to $ \mathcal{B}_b $. Further, one can relate the tetrads $e^n_m$ on $\mathcal{\tilde{\A}}$ to the tetrads $\tilde{e}^\mu_\nu$ on $ \mathcal{B}_b $. To this end, we write
\begin{equation}
\begin{gathered}
\epsilon_{\mu_1 .. \mu_d} (\lambda_{n_1}^{\mu_1} e^{n_1}_{\nu_1} dx^{\nu_1}) \wedge ... \wedge (\lambda_{n_d}^{\mu_d} e^{n_d}_{\nu_d}  dx^{\nu_d}) = \\
= \epsilon_{\mu_1 .. \mu_d} \tilde{e}_{\nu_1}^{\mu_1} ... \tilde{e}_{\nu_d}^{\mu_d} dy^{\nu_1} \wedge ... \wedge dy^{\nu_d} \;.
\end{gathered}
\end{equation}
By comparing the both sides of this expression, we see that 
\begin{equation} 
\tilde{e}^\rho_\mu dy^\mu = \lambda^\rho_n e^n_m dx^m  \;.
\end{equation}
From here, one can read out the metric induced on $ \mathcal{B}_b $ by the bulk metric $g_{nm}$,
\begin{equation}
h_{\mu\nu}=g_{nm}\dfrac{\partial x^n}{\partial y^\mu}\dfrac{\partial x^m}{\partial y^\nu} \; ,
\end{equation}
which coincides with the well--known expression. All other components of the effective theory, including covariant derivatives of matter fields, can be obtained in a similar way.

Now we are ready to generalize Nambu--Goldstone theorem to the case when the dimensionality of the effective theory is smaller than that of the initial theory. Namely, since the stability group $ H_{\mathcal{B}} $ of the effective theory is an intersection of all $ H_b $, in such cases partially broken generators are only those that act trivially on the whole $ \mathcal{B} $. All other generators give rise to physical NG modes, some of which may be gapped. For example, this counting rule shows that for a scalar domain wall the NG mode for the broken Lorentz transformations is redundant. To provide an example of a theory including mNG fields, consider a vector domain wall \cite{Chkareuli:2009ed}. The action of the theory reads
\begin{equation} \label{LagrVect_LongAndTrans}
\mathcal{L} = -\frac{1}{4} F_{mn}^2 - \frac{\lambda}{4} \left( A_m A^m - v^2 \right)^2, ~ F_{mn} = \partial_m A_n - \partial_n A_m \;.
\end{equation} 
The domain wall solution is of the form
\begin{equation} \label{VacVDW}
A_m = \varphi_z \mathfrak{n}_y \;,
\end{equation} 
where $ \mathfrak{n}_y $ is a unit norm vecotr pointing in the $y$ direction.\footnote{The configurations of this type are known to be unstable \cite{Chkareuli:2009ed}. Still, one can use them while studying the Goldstone sector of the theory.} Then, the vector domain wall breaks the following generators,
\begin{equation} \label{PatternDWTranv}
P_z \; , ~~ M_{yi} \; , ~~ M_{zy} \; , ~~ M_{zi} \;.
\end{equation}
However, by applying our criterion for identifying partially broken generators and mNG fields, we see that the NGF for $ M_{zi} $ are redundant while $ \omega_{zy} $, the NGF for the broken Lorentz generator $ M_{zy} $, is massive. This prediction can be straightforwardly verified by a direct computation. 

\bibliography{ref_coset}

\end{document}